\documentclass[runningheads]{llncs}
\usepackage[T1]{fontenc}
\usepackage{graphicx}
\usepackage{verbatim}

\begin{document}
\title{Addressing Trust in AI Systems through Education: A Didactic Perspective}
\titlerunning{Addressing Trust in AI Systems through Education: A Didactic Perspective}
\author{Pierre Haritz\inst{1} \and Hendrik Krone\inst{1} \and Thomas Liebig\inst{1,2}}
\authorrunning{Haritz et al.}
\institute{Department of Computer Science, TU Dortmund University, Dortmund, Germany \and
Lamarr Institute for Machine Learning and Artificial Intelligence, Germany\\
\email{firstname.lastname@tu-dortmund.de}}
\maketitle
\begin{abstract}
Machine learning (ML) education faces two persistent and connected obstacles: many educational tools present ML as an opaque black box, which leaves learners with a superficial understanding, and this same opacity prevents users from forming the calibrated trust that appropriate reliance on AI systems requires. We present ICE-T, a didactic framework that integrates three mutually reinforcing facets: intermodal transfer grounded in Bruner's enactive, iconic, and symbolic modes of representation, computational thinking operationalized through the Use-Modify-Create progression, and explanatory thinking supported by a process model. Connecting the framework to the empirical literature on algorithm aversion, AI literacy, and mental model formation, and to systematic reviews of the K-12 ML activity landscape, we argue that the three facets supply the cognitive mechanisms that the trust calibration literature identifies as drivers of appropriate reliance: representational richness, graduated process control, and the capacity to contextualize errors. On this basis, we propose that trust calibration be treated as an explicit educational objective, with ICE-T as a principled and scalable means of achieving it.
\end{abstract}
\section{Introduction}
\label{sec:intro}

\medskip


The rapid growth of artificial intelligence (AI) and machine learning (ML) technologies in everyday life has prompted widespread calls for their integration into educational curricula at all levels, from primary schools through secondary and higher education~\cite{touretzky2019ai4k12,long2020ailiteracy}. As ML-driven systems increasingly mediate decisions in healthcare, transportation, finance, and social media, the ability to understand, critically evaluate, and ultimately shape these technologies has become a pressing societal concern~\cite{rizvi2023artificial}. The EU AI Act (Art.~3(56)) states that AI literacy denotes ``skills, knowledge and understanding that allow providers, deployers and affected persons, taking into account their respective rights and obligations in the context of this Regulation, to make an informed deployment of AI systems, as well as to gain awareness about the opportunities and risks of AI and possible harm it can cause''~\cite{euaiact2024}. Despite this growing consensus on the importance of AI literacy, how to teach ML effectively to students with little or no prior exposure to programming, in a way that treats trust as a primary concern, remains an open research problem.

Two interrelated issues compound the difficulty of this endeavor. The first is pedagogical: many existing educational tools allow students to interact with ML models at a surface level, for instance by training an image classifier through a web interface, without fostering a deeper understanding of the underlying data pipelines, algorithmic mechanisms, or model evaluation criteria~\cite{wan2019blackbox}. This prevailing treatment of ML as an opaque black box risks leaving students with a superficial and potentially misleading understanding of what ML systems actually do. The second issue concerns trust. Students, and indeed the general public, must develop trust that is neither blindly accepting nor overly dismissive, but calibrated to a system's actual capabilities and limitations~\cite{lee2004trust,glikson2020trust}. These two issues are deeply connected: without understanding how an ML system works, a user has no principled basis for deciding when to rely on it and when to exercise skepticism.

This paper addresses both challenges in an integrated manner and makes three contributions. First, we present the ICE-T concept, a multi-faceted didactic framework for teaching ML that integrates intermodal transfer, computational thinking, and explanatory thinking~\cite{krone2024icet}. Second, we connect this framework to the empirical literature on algorithm aversion, AI literacy, and mental model formation, and show that its facets supply the cognitive mechanisms that the trust calibration literature identifies as prerequisites for appropriate reliance. Third, drawing on systematic reviews of the K-12 ML activity landscape, we show that these mechanisms are systematically undersupported in current practice, and we argue that trust calibration should therefore be treated as an explicit educational objective of ML education.

The remainder of this paper is structured as follows. Section~\ref{sec:preliminaries-on-trust} introduces preliminaries on trust. Section~\ref{sec:foundations} introduces the theoretical foundations of the framework. Section~\ref{sec:tools} reviews the landscape of existing educational tools. Sections~\ref{sec:petsp} and~\ref{sec:icet} present the PETSP-ML process model and the integrated ICE-T concept. Section~\ref{sec:trust} develops the argument connecting ICE-T to trust calibration, and Sect.~\ref{sec:conclusion} concludes.

\section{Preliminaries on Trust}
\label{sec:preliminaries-on-trust}

Historical experience reveals that the relationship between a technology's actual safety and the public's confidence in it is rarely well aligned. In the trust literature, this misalignment is formalized through the concepts of overtrust and undertrust~\cite{lee2004trust,wischnewski2023trustcalibration}. Overtrust arises when users place confidence in a system that does not warrant it. Undertrust, on the other hand, arises when a system that is, in fact, safe and reliable is rejected. Neither extreme serves society well: overtrust exposes individuals to preventable harm, while undertrust forecloses the benefits of genuinely trustworthy systems and can also cause harm in medical applications. The productive middle ground is what Lee and See term \emph{calibrated trust}~\cite{lee2004trust}: a state in which users' confidence in a system tracks its actual capabilities and limitations. Achieving calibrated trust is, in a sense, the common thread running through every historical episode discussed above, in which standards, regulation, and public understanding each contributed, in different eras, to narrowing the gap between what a technology actually was and what the public believed it to be.

\begin{figure}
    \centering
    \includegraphics[width=0.4\linewidth]{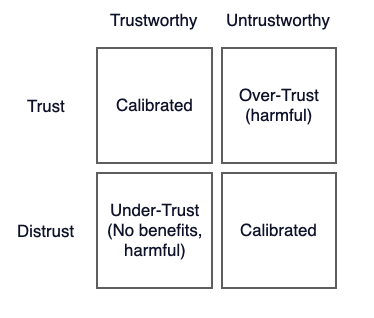}
    \caption{The Trust Matrix depicting the modes of trust towards a system (after~\cite{lee2004trust})}
    \label{fig:trust-matrix}
\end{figure}

Research on human-AI interaction converges on a three-dimensional model of trust formation, in
which trust depends on characteristics of the \emph{trustee} (the AI
system), the \emph{trustor} (the human user), and the \emph{context} in
which they meet~\cite{kaplan2023trustmetaanalysis,li2024developing}.
This framing also exposes why technical work alone is insufficient. A
perfectly safe system that no stakeholder understands does not produce
calibrated trust, but instead produces either blind acceptance (overtrust) or blanket rejection (undertrust),
both of which are failure modes from a deployment perspective (see Figure~\ref{fig:trust-matrix}). Conversely,
communicative effort directed at an unsafe or poorly understood system
produces unjustified confidence. The two dimensions of trustworthiness are therefore coupled: each is a necessary condition
for the usefulness of the other.

Trust in AI is most commonly defined, following~\cite{lee2004trust}, as a user's willingness to be vulnerable to the actions of an automated or intelligent
system in a situation characterized by uncertainty. 
This definition inherits its core logic from interpersonal trust research~\cite{mayer1995integrative}, in which trust is predicated on perceptions of the trustee's \emph{ability}, \emph{benevolence}, and \emph{integrity}. When the trustee is an AI system rather than a human, however, the mapping is imperfect: ability translates
reasonably well into perceptions of functionality and reliability, but benevolence and integrity sit uneasily with non-agentive software, which as prompted extensions to and reconfiguration of the original model~\cite{afroogh2024}.

Two important distinctions structure the contemporary literature. The first separates \emph{cognitive} from \emph{affective} trust: the former rests on rational appraisal of performance and evidence, the latter on emotional responses, identification, and perceived care~\cite{glikson2020trust}. Empirical reviews show that studies of Human-AI interaction typically draw on one or both components, and that the balance between them is itself shaped by cultural context~\cite{aquilino2025decoding}. The second distinction separates \emph{subjective trust} (the user's attitude) from \emph{trustworthiness} (a property of the system itself). Policy frameworks such as the EU's Ethics Guidelines for Trustworthy AI\footnote{\textit{https://digital-strategy.ec.europa.eu/en/library/ethics-guidelines-trustworthy-ai}} operate at the latter level by specifying the conditions under which a system is to be trusted. User-centered research operates at the former, asking when and why people actually do trust.
A well-designed AI system should align the two by producing \emph{calibrated trust} and therefore minimize both overtrust and undertrust.


\section{Theoretical Foundations of Didactics}
\label{sec:foundations}
The didactic approach developed in this work draws on three bodies of educational theory, each of which addresses a distinct but complementary aspect of effective learning in technical domains.

\subsection{Intermodal Transfer: The EIS and Spiral Principles}
\label{subsec:eis}
A foundational element of our framework is Bruner's theory of cognitive representation, which posits that learners encode and process knowledge through three distinct yet interrelated modes: enactive, iconic, and symbolic~\cite{bruner1966studies,bruner1960process}. In the enactive mode, knowledge is acquired and retained through direct physical action and sensorimotor experience. The iconic mode involves the representation of knowledge through mental images, diagrams, and visual models. The symbolic mode, the most abstract of the three, relies on language, mathematical notation, and formal symbol systems.

Bruner argued that effective instruction should guide learners through a progression from concrete, action-based experiences toward increasingly abstract and symbolic forms of understanding. Importantly, these modes are not rigidly sequential stages but coexisting representational systems that can be activated at any point in the learning process. The pedagogical implication is that learners benefit from encountering the same concept through multiple representational modalities, and that the deliberate transfer between these modalities, which we refer to as intermodal transfer, strengthens conceptual understanding and retention.

In the context of ML education, intermodal transfer can be realized, for example, when students first engage in a hands-on classification activity with physical objects (enactive), then represent the same classification task as a visual decision tree diagram (iconic), and finally formalize the classification rules in code or mathematical notation (symbolic). This progression, and the ability to move between modes, is further strengthened by the iterative spiral principle, with the overall goal that abstract ML concepts are grounded in concrete experiences and that students develop the capacity to move flexibly between different levels of abstraction.

\subsection{Computational Thinking and the Use-Modify-Create Framework}
\label{subsec:ct}
The second pillar is computational thinking (CT), a concept that has gained substantial traction in computer science education research as a fundamental skill for all students~\cite{wing2006ct}. CT encompasses a family of cognitive practices, including decomposition, pattern recognition, abstraction, and algorithm design, that enable learners to formulate problems and solutions in ways amenable to computational processing~\cite{grover2013ct,aho2012ct}.

Within our framework, CT is operationalized through the Use-Modify-Create (UMC) progression, a scaffolded approach originally proposed by Lee et al.~\cite{lee2011umc} and subsequently validated in a variety of K-12 educational contexts~\cite{lytle2019umc,houchins2021umc,vieira2023unplugged}. The UMC framework structures learning activities along a gradient of increasing student autonomy and creative ownership. In the Use phase, students interact with pre-built computational artifacts, such as existing ML models or data visualizations, to develop familiarity with their behavior and purpose. In the Modify phase, students make targeted alterations to these artifacts, for instance by adjusting hyperparameters or modifying decision boundaries, and thereby gain insight into how specific changes affect model behavior. In the Create phase, students design and implement their own computational artifacts, which demonstrates a deep and transferable understanding of the underlying concepts.

The UMC progression is particularly well suited to ML education because it mirrors the way practitioners actually work with ML systems. Data scientists rarely build models entirely from scratch; they typically begin by using existing tools and pre-trained models, modify them to fit specific requirements, and only then create novel solutions when the problem demands it. By aligning classroom instruction with this professional workflow, the UMC framework prepares students not only to understand ML concepts in theory but also to engage with them in practice.

\subsection{Explanatory Thinking}
\label{subsec:et}
The third pillar addresses a dimension of learning that is frequently neglected in existing ML education resources: the capacity for explanatory thinking. Explanatory thinking refers to the ability to articulate coherent, logically structured accounts of how and why a system behaves as it does. In the context of ML, this involves understanding not merely that a model produces a particular output given a particular input, but also the role of training data, the mechanics of the learning algorithm, the significance of evaluation metrics, and the potential sources of error or bias. Students are expected to build knowledge through reasoning about science~\cite{kenyon2019explanatory} and are more likely to gain an interest in underlying components~\cite{schwarz2009developing}.

The emphasis on explanatory thinking is motivated by the observation that many current educational tools treat ML as a black box, which allows students to train and deploy models without ever confronting the internal mechanisms that drive model behavior~\cite{krone2024icet}. While such tools can be effective for initial engagement and for demonstrating the practical utility of ML, they fall short of the deeper educational goal, namely to enable students to reason critically about AI systems.

\section{Evaluation of Existing Educational Tools and Platforms}
\label{sec:tools}
To ground our didactic concept in an empirical understanding of the current landscape, we conducted a systematic analysis of prominent educational tools, platforms, and digital games commonly used for teaching ML in school settings~\cite{krone2024icet}. To this end, we selected a representative set of resources and evaluated each one against the three didactic facets: intermodal transfer, computational thinking (UMC), and explanatory thinking.

The evaluation revealed a heterogeneous landscape in which individual tools tend to excel along one or two didactic dimensions but rarely address all three in an integrated manner. Visual programming platforms often support computational thinking through structured coding activities but offer limited opportunities for intermodal transfer or explanatory engagement with the mechanics of ML algorithms. Conversely, web-based ML training tools such as Google Teachable Machine provide compelling iconic and enactive experiences but typically do not expose the algorithmic details that would support explanatory thinking. Very few of the analyzed resources provided a full Use-Modify-Create progression, and most tools limited students to either the Use or the Modify/Create stage without offering meaningful opportunities for guided modification.

This qualitative assessment has since been corroborated at a considerably larger scale. A systematic review of 126 K-12 ML activities, coded according to a taxonomy of algorithmic abstraction that ranges from invisible use to independent creation, confirms that supervised learning activities cluster overwhelmingly in the lowest abstraction levels: invisible use, button interaction, and model deployment together account for almost 60\% of all coded supervised learning entries, while genuinely open-ended creation tasks appear only twice in the entire corpus~\cite{krone2026loa}. A complementary systematic review of 133 activities that analyzed the learning contexts in which ML is introduced reports a similarly narrow picture: a technical and a societal perspective on the same algorithm co-occur in only about three percent of the coded activities~\cite{krone2026calm}. Together, these reviews confirm, with a substantially broader empirical base, the pattern already visible in the smaller tool evaluation: the didactic facets identified in this paper are rarely integrated in existing practice, and algorithmic mechanisms remain largely black-boxed even where use, modification, or contextual framing are already well supported. These findings underscore the need for a holistic didactic concept that explicitly integrates all three facets and guides educators in combining multiple tools and activities.

\section{PETSP-ML: A Process Model for Explanatory Thinking}
\label{sec:petsp}
To promote explanatory thinking in a structured manner, we developed the Promotion of Explanatory Thinking Standard Process for Machine Learning (PETSP-ML)~\cite{krone2024icet}, a process model adapted from the Cross-Industry Standard Process for Data Mining (CRISP-DM)~\cite{wirth2000crispdm}. While CRISP-DM provides a comprehensive and well-validated structure for professional data mining workflows, its original formulation is not directly suited to educational settings, where the objectives, constraints, and prior knowledge of participants differ substantially from those encountered in industrial projects.

\begin{figure}[t]
\centering
\includegraphics[width=0.85\textwidth]{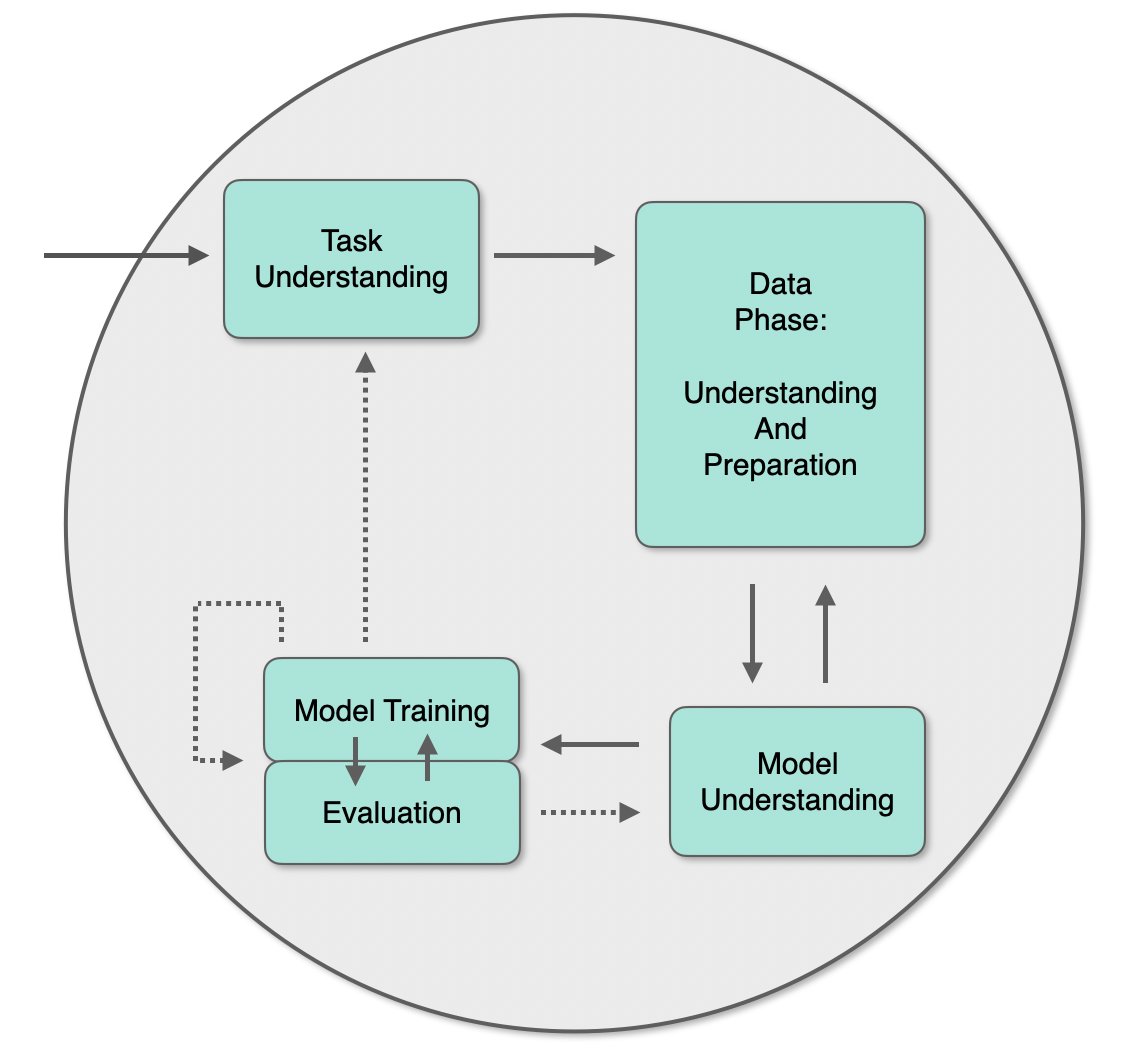}
\caption{PETSP-ML as a process model for teaching ML with a focus on the facet of explanatory thinking. Based on~\cite{krone2024icet}.} \label{fig:petspml}
\end{figure}

PETSP-ML adapts CRISP-DM for the classroom by reorganizing its six phases into four stages aligned with the cognitive and didactic requirements of ML education (see Fig~\ref{fig:petspml})~\cite{krone2024icet}:

\begin{enumerate}
\item \emph{Task Understanding.} Students develop a comprehensive understanding of a given task by actively engaging with its requirements, objectives, and expectations, ensuring clarity and alignment with their learning goals. A central concern at this stage is conveying why ML is an appropriate approach to the problem at hand.
\item \emph{Data Phase.} This stage merges CRISP-DM's Data Understanding and Data Preparation phases. Students collect, inspect, and prepare data for use in training a model, and they engage critically with questions about data quality, representativeness, labeling accuracy, and potential biases.
\item \emph{Model Understanding.} Rather than proceeding immediately to model training, this stage introduces students to the algorithmic principles that underlie the model they will use. It is crucial for gaining insights into how the model makes decisions, understanding its limitations, and ensuring that the model aligns with the intended goals.
\item \emph{Model Training and Evaluation.} Students train their model, evaluate its performance with appropriate metrics, and reflect on the relationship between data characteristics, model design choices, and observed outcomes. The iterative nature of PETSP-ML encourages students to revisit earlier stages, which reinforces the understanding that ML is an iterative, data-dependent process.
\end{enumerate}

The emphasis on model understanding as a distinct, prerequisite stage before model training is a deliberate departure from both professional practice and many existing educational tools, and it is further amplified by the shifted focus away from performance metrics. It reflects the pedagogical priority of building conceptual foundations before procedural skills.

\section{The ICE-T Concept: Integrating the Three Facets}
\label{sec:icet}
The ICE-T concept unifies the three didactic facets described above into a coherent, actionable framework for ML education. The name serves as a mnemonic for its constituent components: Intermodal Transfer (grounded in the EIS principle), Computational Thinking (operationalized through the UMC progression), and Explanatory Thinking (supported by the PETSP-ML process model)~\cite{krone2024icet}.

The core insight of ICE-T is that these three facets are not independent pedagogical strategies to be applied in isolation but mutually reinforcing dimensions that should be woven together throughout a learning unit. Consider a learning unit on decision tree classification of animals based on physical attributes. Students begin with an enactive activity in which they physically sort animal cards according to observable attributes (intermodal transfer: enactive mode). They then transition to constructing a visual decision tree diagram on a whiteboard (iconic mode) and finally to implementing the classification rules in a programming language (symbolic mode). Within each representational modality, students progress through the UMC sequence: they first use a pre-defined sorting game, then modify it by introducing new animal cards or changing the sorting criteria, and finally create their own classification game with original rules. The overall learning unit follows the PETSP-ML process model, from task understanding through data preparation, model understanding, and model training and evaluation.

Importantly, the three facets operate as orthogonal dimensions of a single learning unit, not as a scheme in which every combination of representational mode, UMC stage, and process phase must be traversed. The EIS modes structure how a concept is represented, the UMC progression structures the degree of student autonomy, and PETSP-ML structures the temporal sequence of the unit; which combinations are realized at each point is a deliberate design decision of the educator.

This integrated approach ensures that students encounter ML concepts from multiple cognitive angles, develop both procedural and conceptual knowledge, and are consistently prompted to explain and justify their reasoning.

\section{From Understanding to Trust}
\label{sec:trust}
 Trust in AI is not merely a matter of individual preference; it is a prerequisite for the productive adoption of AI-based tools and, at the same time, a vulnerability that can lead to harmful over-reliance when it is poorly calibrated~\cite{lee2004trust,glikson2020trust}.
\subsection{The Educational Challenge of Calibration}
\label{subsec:calibration}
While the ICE-T concept was originally designed to address the pedagogical challenges of ML education, its implications extend naturally to a second, equally important problem: the formation of appropriate trust in AI systems. As AI systems are deployed in high-stakes contexts, the consequences of both excessive trust (automation bias) and insufficient trust (algorithm aversion) become increasingly severe~\cite{dietvorst2015algorithm,parasuraman2000automation}.

Calibrated trust requires that users hold a sufficiently accurate mental model of the system: they must understand, at least at a functional level, what the system can do, under what conditions it is likely to succeed, and where its failure modes lie. When users lack such a mental model, their trust responses tend to be driven by heuristics, prior attitudes, or superficial cues rather than by a grounded assessment of system behavior~\cite{glikson2020trust}. This observation has profound implications for education, because it suggests that the formation of calibrated trust is fundamentally a learning problem.

The urgency of this problem is amplified by two converging trends. Students frequently use AI tools for writing, problem solving, and information retrieval, yet they often struggle with basic verification strategies and hold significant misconceptions about the accuracy and reliability of AI outputs~\cite{martin2025students,amoozadeh2024trust}. AI literacy shapes trust in AI by influencing both knowledge and perceived value: when faced with a black-box system, literacy tends to reduce trust~\cite{chiang2022exploring}, while at the same time an increased perceived practical value tends to increase reliance~\cite{pan2025ai}. This is particularly relevant because the teaching of machine learning often moves along a spectrum from data-driven to more algorithm-focused perspectives~\cite{MoralesNavarro2024data-algo-spectrum}.

\subsubsection{AI Literacy Recalibrates Rather Than Increases Trust.}
The intuitive view holds that ignorance breeds distrust and that explaining AI to people should therefore raise their trust in it. The empirical record contradicts this in important ways. Chiang and Yin showed that literacy interventions cause users to become more discerning rather than more trusting: participants relied on ML models when the models performed well and withheld trust when outputs were uncertain or flawed~\cite{chiang2022exploring}. Literacy, in their framing, does not increase trust; it recalibrates reliance. A further study showed that digital literacy increased trust in medical AI, but AI literacy specifically reduced it, as deeper awareness of risks generated what the authors call informed skepticism rather than blind confidence~\cite{yao2025trust}. A two-wave panel study across the United States, Spain, and Chile found that AI literacy does not affect trust in AI-generated news directly but works through attitudes, which suggests that literacy interventions need to address both cognition and attitude to shift trust meaningfully~\cite{goyanes2026trust}.

\subsubsection{The Non-Linear Familiarity-Trust Relationship.}
A consistent pattern across the trust literature is that the relationship between familiarity with AI and trust in AI is non-linear~\cite{alruwaili2025modeling}. Novices tend toward algorithm aversion or blanket rejection; users with moderate AI literacy show the most calibrated trust profiles; very high-familiarity power users may over-rely. This inverted-U shape has direct pedagogical implications: the goal of AI education for trust calibration is not to maximize familiarity but to bring users into the calibrated middle range, where they can distinguish between contexts in which the system is likely to succeed and contexts in which it is likely to fail.

Two further findings reinforce this picture. Domain-specific explanations of AI recommendations increased both understandability and willingness to act on those recommendations, while generic explanations produced much weaker effects~\cite{feldman2025impact}. A cross-national teacher survey found that teachers with higher AI self-efficacy and deeper AI understanding reported more calibrated trust, perceiving both more benefits and more concerns, whereas demographic variables had no significant effect~\cite{viberg2025explains}. This positions depth and specificity of conceptual understanding, not exposure or demographics, as main driver for calibrated trust.

\subsection{The Black Box as a Shared Obstacle to Understanding and Trust}
\label{subsec:blackbox}
The pedagogical problem addressed by ICE-T and the trust problem just documented trace back to the same root cause: the treatment of ML systems as opaque black boxes. From a pedagogical perspective, the black-box approach prevents students from constructing the conceptual knowledge needed for deep understanding of ML~\cite{broll2022beyond,ma2023developing}. From a trust perspective, it prevents users from constructing the mental models needed for calibrated reliance. The solution to both problems, we argue, is the same: to open the black box through structured educational experiences that provide multiple complementary perspectives on how ML systems work.

The connection to the algorithm aversion literature is instructive. Dietvorst et al. demonstrated that people who observe an algorithm making predictions are more likely to reject it after witnessing even a single error, because they lack a mental model that would allow them to contextualize the error as an expected consequence of the algorithm's design~\cite{dietvorst2015algorithm}. In a subsequent study, the same authors showed that giving users even a slight degree of control over an algorithm's output, namely the ability to modify its predictions by a small amount, was sufficient to reduce algorithm aversion substantially~\cite{dietvorst2018overcoming}. The key insight is that the feeling of agency and understanding, not the magnitude of control, drove the calibration effect. Cheng and Chouldechova extended this finding by distinguishing between outcome control (the ability to adjust outputs) and process control (the ability to influence how the algorithm works), and showed that both forms of control can mitigate aversion, with process control producing the strongest calibration effects~\cite{cheng2023process}.

This body of research suggests that educational interventions which give students the opportunity not only to observe ML systems from the outside but to step inside them, to enact the algorithm's logic, to modify its parameters, and to observe the downstream consequences, can construct the kind of functional mental model that supports both deep understanding and calibrated trust.

\subsection{Empirical Evidence from Systematic Reviews of Current Practice}
\label{subsec:evidence}
If calibrated trust depends on the depth and specificity of conceptual understanding (Sect.~\ref{subsec:calibration}) and on process control (Sect.~\ref{subsec:blackbox}), then the two ICE-T facets that supply them raise a direct empirical question: \textit{does current K-12 practice actually grant learners the graduated process control that UMC calls for and the explanatory depth that PETSP-ML calls for?} An answer is provided in two systematic reviews of the K-12 ML activity landscape.

\subsubsection{Process Control in Practice.}
A review of 126 activities classifies each one by the role it assigns the learner with respect to the algorithm: the \emph{User} interacts with a model but does not alter its logic, the \emph{Machine} enacts the steps of the algorithm directly, and the \emph{Creator} designs or substantially modifies it~\cite{krone2026loa}. These roles relate to the UMC progression of Sect.~\ref{subsec:ct} without coinciding with it. The User role corresponds to Use and the Creator role to Create, whereas the Machine role, in which learners enact the algorithmic steps themselves, sits closer to the enactive mode of intermodal transfer (Sect.~\ref{subsec:eis}). The intermediate Modify stage remains invisible in this coding.

The distribution of roles is markedly uneven: of 89 coded supervised learning activities, 69 confine the learner to the User role, while only 17 reach Creator and 8 reach Machine. Since even a modest degree of process control substantially reduces algorithm aversion~\cite{dietvorst2018overcoming,cheng2023process}, a landscape in which more than three quarters of supervised learning activities hold learners in the User role leaves this trust-calibrating potential largely untapped. The UMC progression is therefore not a didactic nicety but a corrective to a documented imbalance.

\subsubsection{The Critical-Technical Gap.}
A second, related review examines the contexts in which such activities are framed and the perspective from which learners approach the algorithm, whether as a user, as a technical learner, or as a critical learner who reflects on societal consequences~\cite{krone2026calm,krone2026shifting}. Across a corpus of 133 activities, a technical and a societal perspective on the same algorithm co-occur in roughly 3\% of cases. Model Understanding, the phase in which PETSP-ML asks learners to engage with the internal principles of the algorithm, is reached by a technical perspective in essentially every activity that adopts one, but by a societal perspective in only about 25\%. In the entire corpus, only four activities combine both perspectives.

This critical-technical gap sharpens the argument about explanatory thinking developed in Sect.~\ref{subsec:et}. A learner who observes a biased or unfair outcome can trace it back to identifiable design decisions, such as the properties of the training data or the choice of evaluation criteria, only if the mechanism and its consequences are learned together. Current practice rarely achieves this: where activities examine the mechanism, they typically omit the consequences, and where they examine the consequences, they typically omit the mechanism. Perspective-shifting pedagogy answers this gap. It revisits the same algorithm across the user, technical, and societal perspectives within one coherent sequence, and it thereby extends the spiral principle of Sect.~\ref{subsec:eis} with an explicit critical dimension~\cite{krone2026shifting}.

\subsection{ICE-T as a Framework for Trust Calibration}
\label{subsec:framework}
We argue that the ICE-T concept provides a principled response to these findings, because its three facets address the cognitive mechanisms that the trust calibration literature identifies as the drivers of appropriate reliance.

\subsubsection{Intermodal Transfer and Mental Model Richness.}
The trust calibration literature consistently finds that calibrated trust depends on the depth and specificity of the user's mental model~\cite{lee2004trust,feldman2025impact,viberg2025explains}. The intermodal transfer facet addresses this requirement by ensuring that students encounter each ML concept through multiple representational modalities. A student who has experienced a classification algorithm enactively, iconically, and symbolically is expected to hold a richer and more flexible mental model than a student who has only seen the symbolic formulation, and can draw on different cognitive resources depending on the context.

\subsubsection{UMC and Process Control.}
The UMC progression can be understood as an enactment of the graduated process control mechanism discussed in Sect.~\ref{subsec:blackbox}. In the Use phase, students observe the system from the outside, which the aversion literature predicts will produce fragile trust that collapses at the first observed error. In the Modify phase, students gain the slight degree of control that Dietvorst et al. found sufficient to reduce aversion~\cite{dietvorst2018overcoming}. In the Create phase, students gain the fullest form of process control, the form that produces the strongest calibration effects~\cite{cheng2023process}, by designing the system themselves. The UMC progression thus enacts a graduated increase in the student's agency over the algorithmic process, and the trust calibration literature suggests a corresponding increase in the capacity for calibrated reliance at each step. We note that this argument transfers findings from decision-support settings, in which users rely on a deployed algorithm, to a learning setting in which students build models themselves; the predicted gradient is therefore a testable hypothesis derived from the aversion literature rather than an established result. As the review reported in Sect.~\ref{subsec:evidence} shows, moreover, this graduated exposure is currently the exception rather than the rule.

\subsubsection{Explanatory Thinking and Error Contextualization.}
The core finding of the algorithm aversion literature is that users reject algorithms after observing errors because they cannot contextualize those errors within a functional understanding of the algorithm's design~\cite{dietvorst2015algorithm}. The explanatory thinking facet, operationalized through PETSP-ML, directly addresses this problem by requiring students to construct causal explanations of system behavior. A student who has progressed through the PETSP-ML stages understands not only that a model produces a particular output but why: because the training data had certain properties, because the algorithm partitions the feature space in a particular way, and because the evaluation metric rewards certain behaviors over others. For such a student, an observed error is not an opaque failure of an inscrutable system but a predictable consequence of identifiable design choices and data limitations. This capacity for error contextualization is the cognitive mechanism through which explanatory thinking supports calibrated trust, which fits the capacity that the critical-technical gap shows to be systematically undersupported in current practice.

\subsubsection{The Three Facets as Complementary Trust-Building Mechanisms.}
Taken together, the three facets contribute complementary components to the mental model that calibrated trust requires. Intermodal transfer ensures that students develop a felt sense of what the system does, how it responds to different inputs, and where its outputs seem reliable or unreliable; the enactive grounding provides intuitive understanding, while the iconic and symbolic representations formalize that understanding and make it available for deliberate reasoning. The UMC progression ensures that students discover how the behavior of the system depends on its parameters and design choices. Explanatory thinking ensures that students can assess and articulate whether the design of the system is appropriate for a given context and predict where its failure modes are likely to lie. A mental model that integrates all three components allows the student to make context-sensitive trust decisions: to rely on the system when the conditions align with its design assumptions and to exercise appropriate skepticism when they do not.

\subsection{Trust Calibration as an Explicit Educational Objective}
\label{subsec:objective}
This educational approach is aligned with the empirical evidence that trust calibration is driven by the depth and specificity of conceptual understanding, not by mere exposure or surface-level familiarity. The enactive dimension of intermodal transfer, particularly when students physically enact algorithmic steps, transforms the ML system from an opaque black box into a transparent process whose logic the student has personally internalized. The UMC Create phase further strengthens calibration by revealing the contingency and human authorship of ML systems, which grounds trust assessments in an understanding of design choices rather than in abstract attitudes toward technology. At the same time, the framework addresses the risk that increased AI literacy may lead to over-reliance~\cite{zhang2025exploring}: by exposing students to the full range of an ML system's behavior, including its failure modes and the sensitivity of its performance to design decisions, the Create phase cultivates what the medical AI literature calls informed skepticism~\cite{yao2025trust}, a stance that combines willingness to use AI tools with the critical judgment needed to recognize and respond to their limitations. Moreover, this means that the over-reliance observed among high-familiarity users plausibly stems from extensive exposure without a corresponding mechanistic understanding of failure modes, whereas ICE-T deepens precisely this understanding. The framework thus targets the depth that drives calibration rather than the mere familiarity that drives over-reliance.

School curricula should therefore promote transparency, data literacy, and an understanding of key algorithmic principles. We propose that trust calibration be treated as an explicit educational objective of ML education, alongside traditional objectives such as conceptual understanding and procedural skill. The ICE-T concept provides a concrete mechanism for achieving this objective, and its integration of intermodal transfer, computational thinking, and explanatory thinking ensures that trust calibration is not an isolated add-on but an organic consequence of a well-designed, multi-faceted learning experience.

\section{Conclusion}
\label{sec:conclusion}
This paper presented an integrated approach to ML education that addresses both the challenge of effective pedagogy and the challenge of trust calibration. The ICE-T concept combines intermodal transfer based on Bruner's EIS principle, computational thinking operationalized through the Use-Modify-Create progression, and explanatory thinking supported by the PETSP-ML process model~\cite{krone2024icet}. An evaluation of existing educational tools, corroborated by systematic reviews of 126 and 133 K-12 ML activities~\cite{krone2026loa,krone2026calm}, identified significant gaps in current practice: the widespread treatment of ML as a black box, the confinement of learners to passive user roles, and the near-complete separation of technical from societal perspectives. Connecting these findings to the empirical literature on algorithm aversion and AI literacy~\cite{dietvorst2015algorithm,chiang2022exploring,yao2025trust}, we argued that the three ICE-T facets supply exactly the cognitive mechanisms that calibrated trust requires: representational richness, graduated process control, and the capacity to contextualize errors. On this basis, we proposed treating trust calibration as an explicit educational objective of ML education.

The central limitation of this work is that the connection between ICE-T and trust calibration is, at this stage, theoretical: it derives testable predictions from the empirical trust literature but has not itself been evaluated, and its empirical grounding is limited to the K-12 landscape. Future work should therefore assess ICE-T-based learning units in classroom studies using validated measures of trust calibration, examine whether the predicted gradient of the UMC progression manifests in students' reliance behavior, and investigate how the framework transfers across age groups and ML paradigms beyond supervised learning.

%
\bibliographystyle{splncs04}
\bibliography{references}
\end{document}